\documentclass[useAMS]{mn2e}
\usepackage{times}
\usepackage{amssymb}
\usepackage{rotating}
\input{epsf}

\title[Jets and the Accretion Flow in Low Luminosity Black Holes]
{Jets and the Accretion Flow in Low Luminosity Black Holes}

\author[E. Gardner and C. Done]
{Emma Gardner and Chris Done\\
Department of Physics, University of Durham, South Road,
Durham DH1 3LE, UK\\}

\date{Submitted to MNRAS}
\pagerange{\pageref{firstpage}--\pageref{lastpage}} \pubyear{}

\def\apj{ApJ}

\begin{document}

\topmargin = -0.5cm

\maketitle

\label{firstpage}

\begin{abstract}

The X-ray spectra of black hole binaries in the low/hard state first
harden as the flux decreases, then soften. This change in behaviour
has been variously attributed to either the X-rays switching from
being produced in the flow to being dominated by the jet, or to the
flow switching seed photons from the disc to self generated seed
photons from cyclo-synchrotron.  Here we build a simple truncated
disc, hot inner flow, plus standard conical synchrotron jet model to
explore what this predicts for the X-ray emission mechanism as a
function of mass accretion rate.

We find that the change in X-ray spectral index can be quantitatively
(not just qualitatively) explained by the seed photon switch in the
hot flow i.e. this supports models where the X-rays are always
produced by the hot flow.  By contrast, standard conical jet models are
as radiatively inefficient as the hot flow so there is no transition
in X-ray production mechanism with $\dot{m}$. Including the effects of electron cooling allows the jet X-rays to drop more slowly with accretion rate and hence overtake the X-rays from the hot flow, however this produces a corresponding change in the radio-X-ray
correlation, which is not observed. We argue that the unbroken
radio-X-ray correlation down to quiescence rules out the jet transition model as an explanation for the trend in X-ray spectral index.

Our favoured model is then a truncated disc with an inner, hot,
radiatively inefficient flow which always dominates the hard X-rays,
coupled to a conical synchrotron jet which produces the radio
emission. However, even this has issues at low $\dot{m}$ as the low
optical depth and high temperature of the flow means that the Compton
spectrum is not well approximated by a power law. This shows the need
for a more sophisticated model for the electron distribution in the hot
flow. 

\end{abstract}


\section{Introduction} \label{sec:introduction}

The low/hard state (LHS) of black hole binaries (BHB) is 
typically seen at mass accretion rates below a few per cent of the 
Eddington limit. It is characterised by a hard X-ray spectrum, rising
in $\nu f_\nu$ to a peak at a few hundred keV, in sharp contrast to
the typical temperature of a few hundred eV expected from an
optically thick, geometrically thin accretion disc. These X-rays are
also strongly variable on short (sub second) 
timescales, again, in sharp contrast to the long (few hour) 
viscous timescale expected from even the innermost radii of a
thin disc. These properties instead are more typical of the 
alternative set of solutions of the accretion flow equations, where
the flow is geometrically thick, and optically thin. The most
well known of these alternative solutions are the Advection Dominated
Accretion Flow (ADAF) models (Narayan \& Yi 1995), but these
are only an analytic approximation to what is almost certainly a
more complex solution, as the flow must be threaded by magnetic 
fields. Differential rotation shears the field azimuthally, 
while buoyancy lifts it vertically, and the combination sets up a 
turbulent magnetic dynamo which acts to transport angular momentum
outwards so material can fall inwards. Close to the horizon, 
this turbulent field can also produce a jet, as observed in this state
(see e.g. the review by Done, Gierlinski \& Kubota 2007 hereafter 
DGK07, Fender 2004).

These hot flow solutions are only possible at low mass accretion rates, 
collapsing to the standard disc solutions when the flow becomes 
optically thick. This gives a mechanism for the dramatic hard to soft state
transition seen in the BHB, and the associated collapse of the radio jet.
This transition is complex, but the data can be largely fit into a picture
where the thin disc progressively replaces the hot flow down to smaller
radii as the mass accretion rate increases. These truncated disc models
predict that the contribution from the thin disc becomes stronger with
increasing mass accretion rate, increasing the seed photons for
Compton cooling of the hot flow, so the hard X-ray spectrum becomes
softer, as observed. All timescales associated with the 
disc truncation radius will decrease, giving a qualitative (and now quantitative)
framework in which to explain the correlated increase in characteristic
frequencies of the time variability (DGK07, Ingram, Done \& Fragile 2009; 
Ingram \& Done 2011; 2012). 

While this is an attractive picture, it is still somewhat controversial. 
The LHS sometimes shows a soft component whose temperature and
luminosity imply a very small emitting area, not consistent with the
large radius expected for a truncated disc 
(Rykoff et al 2007; Reis et al 2010). 
While some of this can be explained by continuum modelling,
irradiation and assumptions about the inner disc boundary condition
(e.g. Gierlinski, Done \& Page 2008; Makishima et al 2008), there
is still an issue for the lowest mass accretion rate spectra
(Reis, Miller \& Fabian 2009). However, this can still be consistent
with the truncated disc picture if this 
component is instead produced by clumps
torn from the edge of the disc in the truncation process 
(Chiang et al 2010), which might also explain the variability seen in this 
component (Uttley et al 2011). The radii derived from the iron line
profile are even more controversial, but issues with
instrumental effects and continuum modelling again mean that this
is not definitive evidence ruling out a truncated disc for the LHS
(c.f. Miller et al 2006; Done \& Diaz Trigo 2010; Reis et al 2010; 
Kolehmainen, Done \& Diaz Trigo 2011). Thus we assume a 
truncated disc geometry in this paper, and quantitatively explore
its consequences for the emission spectra of BHB. 

In particular, the data show that the X-ray spectral index first
becomes harder, but then softens again below an (Eddington scaled)
luminosity of $L/L_{Edd}\sim 10^{-2}$ (e.g. Corbel, Koerding \& Kaaret
2008; Russell et al 2010; Sobolewska et al 2011 hereafter S11). The
truncated disc/hot inner flow model gives a possible explanation for
this behaviour of the X-ray spectral index. The disc recedes as
mass accretion rate drops, which leads to a decrease in the seed
photon luminosity intercepted by the hot flow so its Compton spectrum
hardens. However, there is another source of seed photons, 
from cyclo-synchrotron emission generated within the flow itself by
the hot electrons spiralling in the turbulent magnetic field. This
source of seed photons increases as the mass accretion rate drops, as
the drop in density means that the emission is much less self-absorbed
and this more than compensates for the drop in emissivity. Thus the
flow should make a transition from hardening due to Compton scattering
on the receding disc, to softening due to Compton scattering of
cyclo-synchrotron photons within the flow (S11). 

While this works qualitatively, S11 did not test whether this could
work quantitatively. Here we build a simple truncated disc/hot inner
flow model and explore whether this can indeed match the observed
X-ray spectral index behaviour. Such models have been built before but
these often focus on the ADAF, so neglect seed photons from an outer
truncated disc (Merloni, Heinz \& di Matteo 2003; Yuan, Cui \& Narayan
2005; Yuan et al 2007; Qiao \& Liu 2013). The ones which do include seed photons from
the disc use such a large disc truncation radius ($10^4$ Schwarzschild
radii), that these have negligible effect (Narayan, Barret \&
McClintock 1997, Esin, McClintock \& Narayan 1997).

We also explore the alternative possibility for the change in X-ray
spectral index, where this marks instead the change from a flow
dominated to a jet dominated X-ray spectrum (Russell et al 2010;
S11). Such a transition to a jet dominated flow (JDAF) was predicted
by Yuan \& Cui (2005) and there are coupled ADAF-jet
models in the literature where the X-rays are produced by synchrotron
jet emission at low mass accretion rates (e.g. Yuan \& Cui 2005; Yu,
Yuan \& Ho 2011).

Yet another approach to explain the low/hard state uses JDAF models at
all $\dot{m}$, so that the X-rays are always dominated by the jet
(e.g. Falcke, Koerding \& Markoff 2004). However,
more recent JDAF models now also include some (often dominant)
contribution to the hard X-ray spectrum from Comptonisation in a hot
flow (Markoff et al 2005; Nowak et al 2011). The focus in these papers
has been to (successfully) fit the observed SED's, rather than
systematically exploring the predicted behaviour of the model as a
function of $\dot{m}$ for physically motivated scalings, which is the
aim of this work.

Here we make a simple model of the accretion flow (truncated disc and
hot, radiatively inefficient inner flow) to gain a quantitative
understanding of how the X-ray spectrum evolves with $\dot{m}$ in
terms of the contribution of disc and cyclo-synchrotron seed photons
for flow Comptonisation.  We then couple this to a standard conical
jet model (Blandford \& Konigl 1979; Merloni et al 2003) to assess the
relative contribution of the flow and the jet to the X-ray
emission. We use independent constraints on the relative contribution
of flow and jet from the observed radio-X-ray correlation
(Hannikainen et al. 1998; Corbel et al 2003; Gallo, Fender \& Pooley
2003; with a more recent compilation in Corbel et al 2013).

We show that a truncated disc and hot inner flow model can
quantitatively as well as qualitatively explain the observed change in
behaviour of the hard X-ray spectral slope at lower luminosities. By
contrast, it is very difficult to make a model in which the X-rays
switch from being produced in the flow to being produced in the
synchrotron jet. Standard conical jet models are as radiatively
inefficient as the hot flow, as both magnetic energy density and
relativistic particle pressure both scale as $\dot{m}$, so the
synchrotron radiation (which is their product) scales as $\dot{m}^2$
(e.g. Merloni et al 2003; Falcke et al 2004).  Changing the jet
scalings to force such a switch produces a clear break in the
radio-X-ray correlation (see also Yuan \& Cui 2005), which is not
observed.

Our favoured model is then a truncated disc with a hot,
radiatively inefficient inner flow which always dominates the hard X-rays,
coupled to a conical synchrotron jet which produces the radio
emission. However, even this has problems at low $\dot{m}$ as the low
optical depth and high temperature of the flow means that the Compton
spectrum is not well approximated by a power law, subtly distorting
the radio-X-ray correlation from the observed $L_{R}\propto
L_X^{0.7}$ relation. This shows the need for a more sophisticated
accretion flow model, perhaps including non-thermal electrons (Malzac
\& Belmont 2009; Vurm \& Poutanen 2009), and/or an inhomogeneous flow
(Veledina, Poutanen \& Vurm 2012)

\section{The fiducial truncated disc/hot inner flow model}

In all the following we use dimensionless 
radii $r=R/R_g$ where $R_g=GM/c^2$,
and mass accretion rates $\dot{m}=\dot{M}/\dot{M}_{Edd}$, where the
Eddington limit $L_{Edd}=\eta\dot{M}_{Edd}c^2$ and $\eta=0.057$ for a
Schwarzschild black hole with innermost stable
circular orbit $r_{isco}=6$. We plot models for a $10 M_{\odot}$ black
hole. 

Our main aim is to explore the origin of the X-ray flux in the
low/hard state, firstly whether the truncated disc/hot inner flow
model can produce the observed change in behaviour of spectral index
with $\dot{m}$, and then to see whether this can also be produced by
jet models. Previous models which included both truncated disc seed
photons and internally generated cyclo-synchrotron seed photons
(Narayan et al 1997; Esin et al 1997) did not explicitly explore
this, and are also based on a pure ADAF model for the accretion flow.
Such pure ADAF models are too hot and optically thin (maximum optical
depth $\tau\propto \dot{m} <1$) to match the observed hard X-ray
emission (Yuan \& Zdziarski 2004). Allowing advection to be negative
(heating the flow) as well as positive (assumed in the ADAF solution)
takes the (luminous hot accretion flow: LHAF) models closer to the
data, but there is still a clear mismatch, with data extending up to
an optical depth of $\tau\sim 2$ (Yuan \& Zdziarski 2004). This
probably reflects the fact that all such analytic models are only an
approximation to a more complex reality, with magnetic fields
threading the flow. Hence rather than build a full ADAF/LHAF model,
which is known to not match the data, we instead take the key
aspects of these models (radiatively inefficient flow i.e. luminosity
$L\propto \dot{m}^2$, which exists only up to a maximum mass accretion
rate, $\dot{m}_{c}$) and set the parameters of this flow from the data
i.e. we take $\dot{m}_{c}=0.1$ and $\tau=\tau_{max} (\dot{m}/\dot{m}_c)$
with $\tau_{max}=2$ (e.g. Ibragimov et al 2005; 
Torii et al 2011; Yamada et al 2013).

In the truncated disc/hot inner flow geometry, this
radiatively inefficient flow exists inside a Shakura-Sunyaev disc
truncated at radius $r_{t}\ge r_{isco}$.  Evaporation of the cool
disc by thermal conduction from a hot corona is known to produce this
geometry at low mass accretion rates (Liu et al 2002; Mayer \& Pringle
2007), where it typically gives $r_{t}\propto \dot{m}^{-1/2}$
below the critical mass accretion rate, $\dot{m}_{c}$, at which the
hot flow collapses (e.g. Czerny et al 2004). Evaporation models show
that the disc is still substantially truncated at this critical mass
accretion rate, but the value of this minimum truncation radius is
$\sim 40 R_g$ ($20R_{sch}$: Czerny et al 2004). However, the
evaporation rates assume the hot flow is an ADAF, whereas our flow is
denser and cooler. The conductive flux depends more strongly on
density, so we expect stronger evaporation. This combined with weaker
constraints on the observed disc radius in the low/hard state (Yamada
et al 2013) motivates us to choose $r_{t}=20
(\dot{m}/\dot{m}_c)^{-1/2}$.

We assume a standard Novikov-Thorne emissivity for a disc from
$r_{out}=10^5$ to $r_{t}$, and assume that all this energy
thermalises, giving $L_{disc}$.  The remaining energy from the
Novikov-Thorne emissivity from $r_{t}$ to $r_{isco}$ is available
to power the hot flow, $L_{hot,power}$, but this is radiatively inefficient so we take the actual radiated power to be $L_{hot}=(\dot{m}/\dot{m}_{c})
L_{hot,power}$ i.e. assume that the flow is as efficient as a thin
disc at $\dot{m}_c$.

The hot flow radiates $L_{hot}$ via Comptonisation (which depends on
seed photon luminosity from both the disc and cyclo-synchrotron
photons generated by the electrons interacting with the magnetic field
in the hot flow) and bremsstrahlung (which depends on density).  We
assume that the hot flow is a homogeneous sphere. The obvious radius
of this sphere is $r_{t}$, but the emission should be centrally
concentrated, so instead we assume that all the energy is dissipated
in a region $r_h=20$.  At any radius $r$ in the disc, we calculate the
fraction of photons illuminating the hot flow, so the seed photon
luminosity, $L_{seed,disc}$ is given by this integrated over all the
disc from $r_{out}$ to $r_{t}$.  The density of the flow is then
$n\sim \tau/(\sigma_T r_h R_g)$.

Radiatively inefficient flows are also generically two temperature,
with ion temperature set by the virial temperature $kT_{ion}\approx
m_p c^2/r$, while the electron temperature is set by the balance of
heating and cooling.  We assume that the flow is homogeneous within
$r_h$ so $kT_{ion}\sim m_p c^2/r_h$. Simulations show that the energy
density in the tangled magnetic field saturates to $\sim 10$ per cent
of the gas pressure, so $U_B=B^2/(8\pi)=0.1 nkT_{ion}$.  The
cyclo-synchrotron emission from the hot flow then extends as an
approximate steep power law from $v_B=eB/(2\pi m_e c)=2.6\times 10^6
B$. However, the majority of this emission is self-absorbed, so the
emission peaks instead at the self-absorption frequency
$\nu_{csa}=\frac{3}{2}\nu_B\theta_e^2 x_m$ where the electron
temperature $\theta_e=kT_e/m_e c^2$ (found iteratively, see below) and
$x_m$ typically has values of a few hundred to a few thousand (see
Appendix for full details). The luminosity is then
$L_{seed,cyclo}\propto n\nu^2_{csa} V$ where $V=\frac{4}{3}\pi r_h^3
R_g^3$ is the volume of the hot flow.

The total seed photon luminosity
$L_{seed}=L_{seed,disc}+L_{seed,cyclo}$. We take the seed photon
energy ($\nu_{seed}$) as the weighted mean of the inner disc
temperature and the cyclo-synchrotron self-absorption frequency, as
defined in equation A11 in the Appendix. The electron temperature can
then be derived self consistently from balancing heating ($L_{hot}$)
and cooling (determined by $L_{seed}$, but also including
bremsstrahlung) rates using the publicly available {\sc eqpair}
code. This calculates the electron temperature and resulting emission
spectrum from a homogeneous sphere, given inputs of the heating power
to the electrons ($L_{hot}$), the optical depth and size of the region
($\tau$ and $r_h$), and the power and typical energy of the seed
photons ($L_{seed}$ and $\nu_{seed}$) for Compton cooling (Coppi
1999).  The resulting spectrum incorporates both bremsstrahlung and
Compton components and does not assume that the Compton emission can
be approximated as a power law.  This is increasingly important as the
flow density drops, as each successive Compton order scattering is
separated by a factor $1/\tau$, making the spectrum increasingly bumpy
as the mass accretion rate decreases.

To summarise: our accretion flow model consists of a truncated disc
where the truncation radius increases with decreasing $\dot{m}$, and a
radiatively inefficient inner hot flow powered by the remaining
gravitational energy that is not dissipated in the truncated disc. We
allow the optical depth of this hot flow to decrease with $\dot{m}$,
and use both the external disc photons intercepting the hot flow and
internal cyclo-synchrotron photons generated within the hot flow as
seed photons for Comptonisation.
 
\begin{figure} 
\centering
\begin{tabular}{l}
\leavevmode  
\epsfxsize=8cm \epsfbox{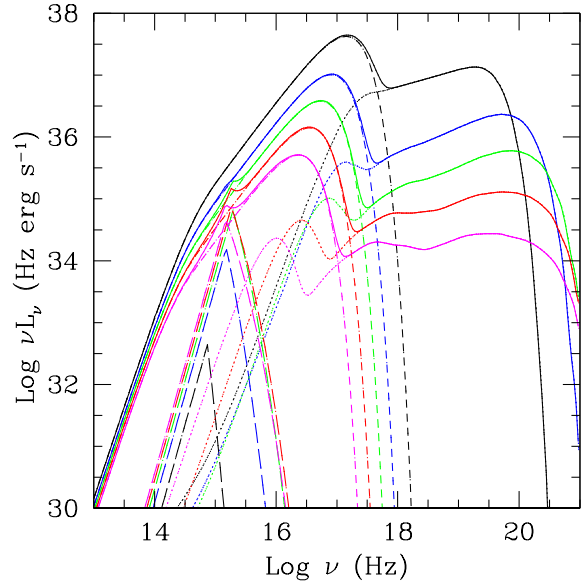}\\ 
\end{tabular}
\caption{Model SEDs, with truncated
disc (short dashed line), hot flow cyclo-synchrotron emission (long
dashed line) and Comptonisation of both disc and cyclo-synchrotron
seed photons (dotted line) for increasing truncation radius: $20R_g$
(black), $35R_g$ (blue), $50R_g$ (green), $70R_g$ (red) and $100R_g$
(magenta). Solid line shows sum of all three components.}
\label{fig1}
\end{figure}

\begin{figure*} 
\centering
\begin{tabular}{l|c|r}
\leavevmode 
\epsfxsize=5.5cm \epsfbox{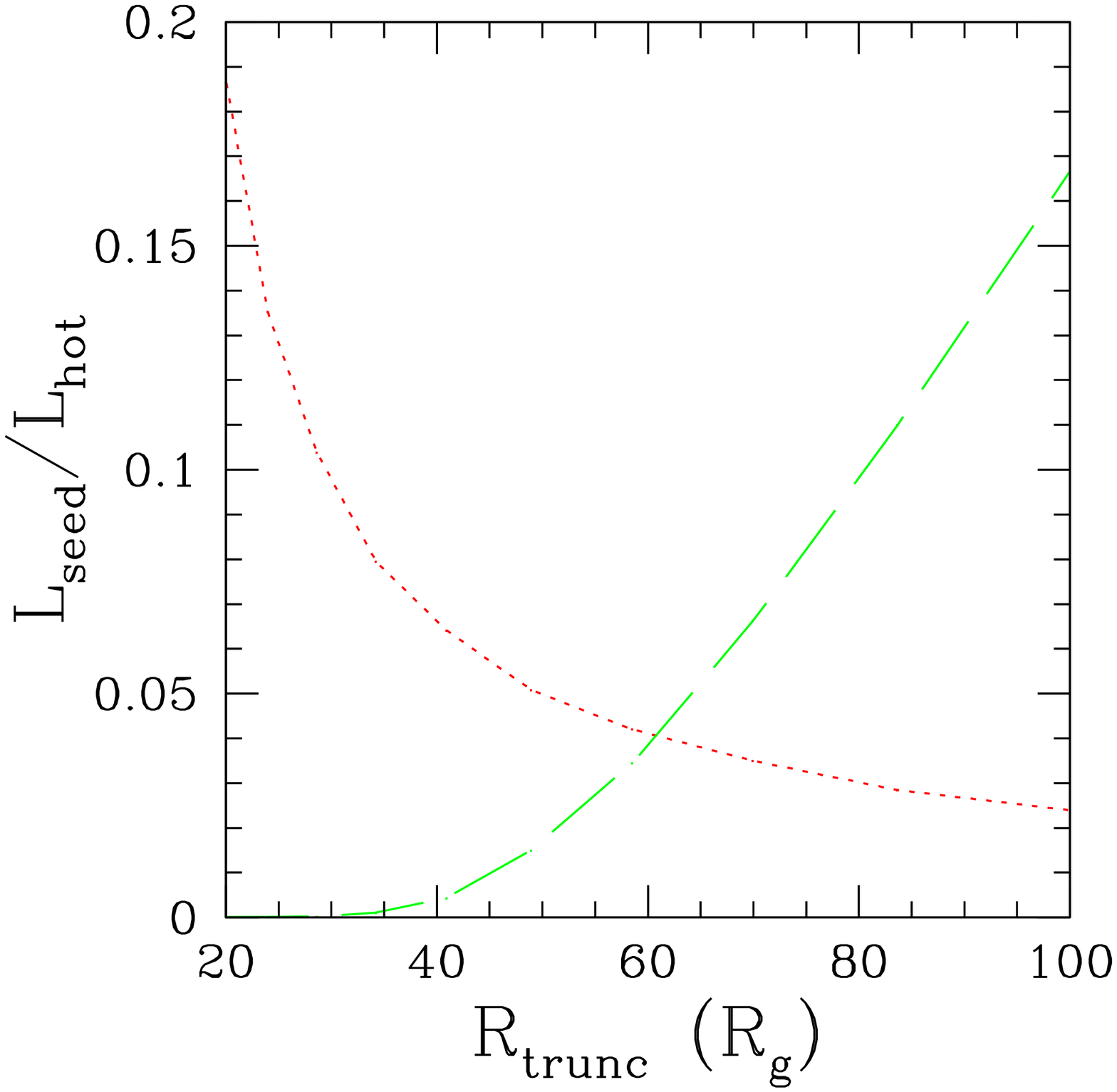} &
\epsfxsize=5.5cm \epsfbox{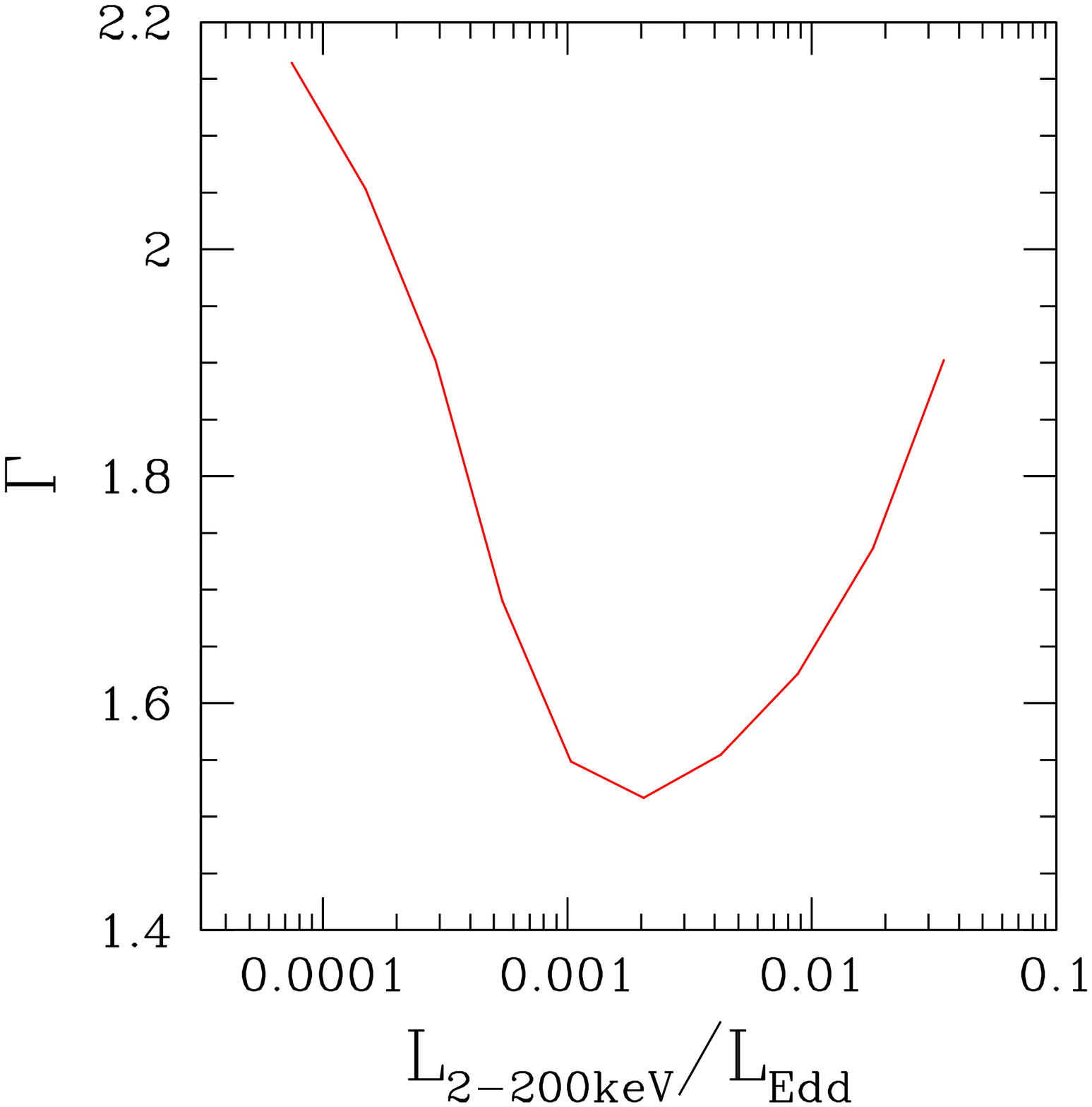} &
\epsfxsize=5.5cm \epsfbox{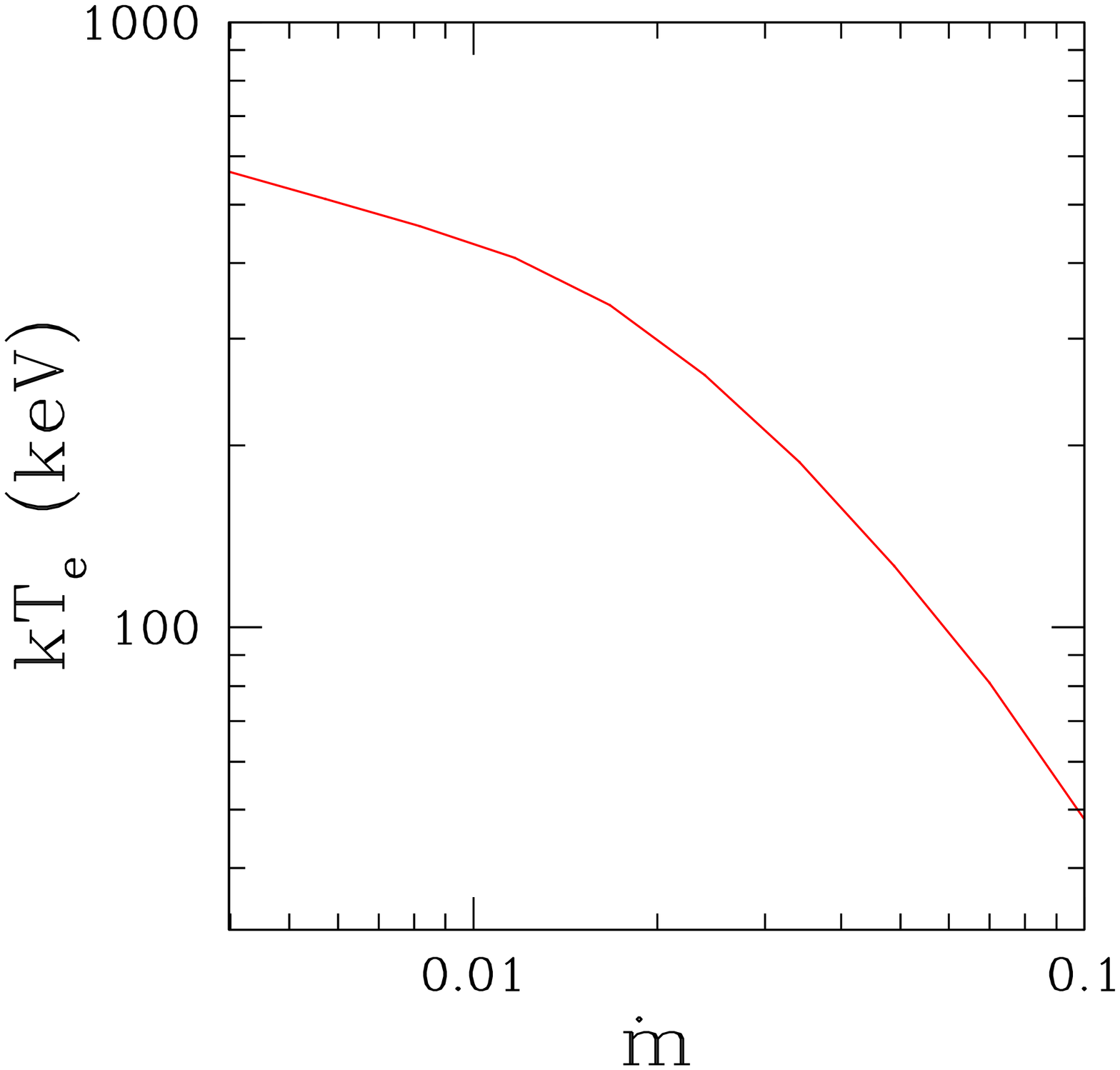}\\
\end{tabular}
\caption{a). Seed photon luminosity as a function of truncation radius, for disc seed photons (red dotted line) and cyclo-synchrotron seed photons (green dashed line). For $R_{trunc}\gtrsim 60R_g$ ($\dot{m}\lesssim 0.01$) the dominant source of seed photons is cyclo-synchrotron emission from the hot flow. b). Photon index as a function of 2-200keV X-ray luminosity, showing softening of the X-ray spectrum at low luminosities as cyclo-synchrotron seed photons begin to dominate. c). Hot flow electron temperature as a function of mass accretion rate, where $\dot{m}=\dot{M}/\dot{M}_E$.}
\label{fig2}
\end{figure*}

\subsection{Spectral Changes with Accretion Rate}

Fig 1 shows a sequence of model spectra for
$\dot{m}=\dot{m}_c=10^{-1}$ to $\dot{m}=4\times 10^{-3}$
(i.e. $r_t=20$ to $100$). Solid lines show the total emission, long
dashed, short dashed and dotted lines show the individual components
of cyclo-synchrotron, truncated disc and Comptonisation, respectively.

The proportion of luminosity in the disc compared to the hot flow,
$L_{disc}/L_{hot}\approx (\dot{m}/\dot{m}_c)^{-1}
(\frac{r_t}{r_{isco}}-1)^{-1}$. Since we also know how $r_t$ depends
on $\dot{m}$ we can simplify this further to $\approx 0.3
(\dot{m}/\dot{m}_c)^{-1/2}\propto r_t$. Thus decreasing $\dot{m}$ by a
factor of 25 increases the disc truncation radius by a factor of 5 and
increases $L_{disc}/L_{hot}$ by a factor 5. This is not a large
factor, but is evident in Fig 1 by comparing the ratio between the
peak $\nu f_\nu$ flux of the disc and Comptonised emission for the
highest and lowest $\dot{m}$ spectra.

However, the ratio between $L_{seed,disc}/L_{hot}$ changes by much
more than $L_{disc}/L_{hot}$ as the fraction of seed photons
intercepted by the hot flow drops as $r_t$ increases.  The seed
photons from the disc which illuminate the hot flow are integrated
over the entire disc, but both the disc luminosity and the fraction
which are intercepted by the flow will peak at $r_t$. Hence
$L_{seed,disc}\approx L_{disk} (r_h/r_t) \arcsin(r_h/r_t) \approx
\frac{\dot{m}}{r_t} (r_h/r_t)^2 \propto \dot{r_t}^{-5}\propto
\dot{m}^{2.5}$.  Thus $L_{seed,disc}/L_{hot}\propto \dot{m}^{2.5}
/\dot{m}^2 \propto \dot{m}^{1/2}\propto r_t^{-1}$. Thus while
$L_{disc}/L_{hot}$ increases by a factor 5 as $\dot{m}$ decreases,
$L_{seed,disc}/L_{hot}$ decreases by a factor 5 (see red line in Fig
2a). If this were the only source of seed photons, the spectrum
should harden substantially. However, there are also seed photons from
the cyclo-synchrotron emission. These have $L_{seed,cyc}\propto n
\nu_{csa}^2 \propto n (B \theta_e^2)^2 \propto \dot{m}^2\theta_e^4$
, so $L_{seed,cyc}/L_{hot}\propto \theta_e^4 $.  This increases 
as $\dot{m}$ decreases, as $\theta_e$ increases as accretion rate 
drops (see below). 
The green line on Fig 2a shows the internally generated cyclo-synchrotron emission starts to dominate over seed photons from the disc 
at $r_t>60$ (equivalently $\dot{m}\le 10^{-2}$). Thus the total $L_{
seed}/L_{hot}$ reaches a minimum at this point, and then starts to 
increase. This change in dominant seed photons can also be seen in 
Fig 1 as the Compton spectrum extends to lower energies reflecting
the lower seed photon energy of the cyclo-synchrotron photons.

The Comptonisation spectral slope is set by $L_{seed}/L_{hot}$, so
this also shows a minimum corresponding to the minimum
$L_{seed}/L_{hot}$. Fig 2b shows the resulting 2-10~keV power law
index and 2-200~keV bolometric luminosity, $L_{2-200}$. 
Our minimum in photon index occurs $2-3\times 10^{-3} L_{Edd}$, a 
factor 2-3 below that shown by the data in S11. Given the simple assumptions made about the structure of the flow, this is probably
not significant. In particular, changing the efficiency of the 
hot flow from a simple $\propto \dot{m}$ to the more complex
behaviour calculated by 
Xie \& Yuan (2012) specifically for an ADAF model would 
make this discrepancy smaller. 
Thus the model is able to
quantitatively describe a key observation of the low/hard state, namely
that the X-ray spectrum hardens with decreasing $\dot{m}$ and then
softens again by the change in seed photons from the disc to
internally generated cyclo-synchrotron. This softening of the Comptonised
emission can be seen by eye in the spectra of Fig 1 by comparing
the slope of the tail at highest and lowest luminosity.

Qiao \& Liu (2013) find a similar trend in photon index using a full ADAF calculation. However their disc-corona geometry is rather different from that considered here. They focus on 
the residual inner disc which can remain after considering
thermal conduction from the hot flow (Liu et al 2006; Liu, Done \& Taam 2011). At the highest mass accretion
rates, the disc extends all the way down to the ISCO. Then as accretion rate drops 
a gap opens up between the inner and outer disc at $\sim 200R_g$, and this
gap extends inwards and outwards as $\dot{m}$ decreases until the entire inner disc evaporates. Thus their drop in seed photons comes
from a decreasing outer extent of the inner disc, whereas in our model it comes from the increasing inner radius of the outer disc. 
Nonetheless, both models have a drop in seed photons with mass accretion rate, so the spectra harden, and then both models
show the characteristic minimum as self generated cyclo-synchrotron photons take over as the dominant seed photons in the hot flow. 

Fig 2c shows the resulting electron temperature, set from the balance
of heating and cooling. The heating rate is $\propto \dot{m}^2$, and
cooling is predominantly Compton cooling so is $\propto 4\theta_e\tau
L_{seed}$.  At high $\dot{m}$, the seed photons are from the disc so
the cooling rate is $\propto 4\theta_e\tau \frac{\dot{m}}{r_t}
(\frac{r_h}{r_t})^2$.  Hence $\theta\propto \dot{m}^{-3/2}$, quite
close to the observed dependence. Conversely, when seed photons from
cyclo-synchrotron cooling dominate, the Compton cooling rate is
$\propto 4\theta_e\tau n\nu_{csa}^2$, where $\nu_{csa}\propto B
\theta_e^2$ so $\theta_e \propto \dot{m}^{-0.2}$. The strong increase
in seed photons with increasing temperature leads to increasing
cooling with decreasing mass accretion rate, which counteracts much of
the decrease in cooling from the decrease in optical depth. Thus the
electron temperature increases much more slowly as the mass accretion
rate decreases. Again this can be seen in the spectra of Fig 1, where
the electron temperature (marked by the high energy rollover of the
tail) first increases markedly with decreasing mass accretion rate,
then stabilises. This changing temperature dependence on accretion
rate is a testable prediction of the model. Current observations
already show that the behaviour in the brightest low/hard states do
indeed show the predicted decrease in temperature with increasing
$\dot{m}$ (Motta et al 2009; Torii et al 2011), but future
observations with the more sensitive Soft Gamma Ray detector
(60-600~keV bandpass) on ASTRO-H (Takahashi et al 2012) will be able
to constrain the temperature down to much lower $\dot{m}$.

The other obvious change in the Comptonised emission is that it is
progressively less well described by a power law as $\dot{m}$
decreases and $\tau\ll 1$.  At such low optical depths the individual
scattering orders become visible, giving a more complex spectral
shape. Data are rarely fit with such low optical depths, as X-ray
observations do not show the strong first Compton peak, which would be
clearly visible in the X-ray regime if the seed photons were provided
by a disc. From our model it is clear the dominant source of seed
photons at these mass accretion rates is cyclo-synchrotron
emission. This brings the first peak out of the X-ray regime, leaving
the X-ray spectrum to be dominated by higher order scattering with
less extreme curvature. Nevertheless, this is still not visible in
X-ray spectra from low $L/L_{Edd}$ flows (e.g. Corbel et al 2006).  We
suggest the reason for this is that our model assumes that the
electrons in the hot flow completely thermalise. An initially
non-thermal acceleration process will probably thermalise via
self-absorption of its own cyclo-synchrotron radiation in bright
low/hard states (Malzac \& Belmont 2009; Poutanen \& Vurm
2009). However, the thermalisation timescale increases as the source
luminosity drops, so the electron distribution retains more of its
initially non-thermal character, giving a non-thermal power law Compton spectrum (e.g. Veledina et al 2011). 

\section{Fiducial conical jet}

\begin{figure*} 
\centering
\begin{tabular}{l|c|r}
\leavevmode  
\epsfxsize=5.5cm \epsfbox{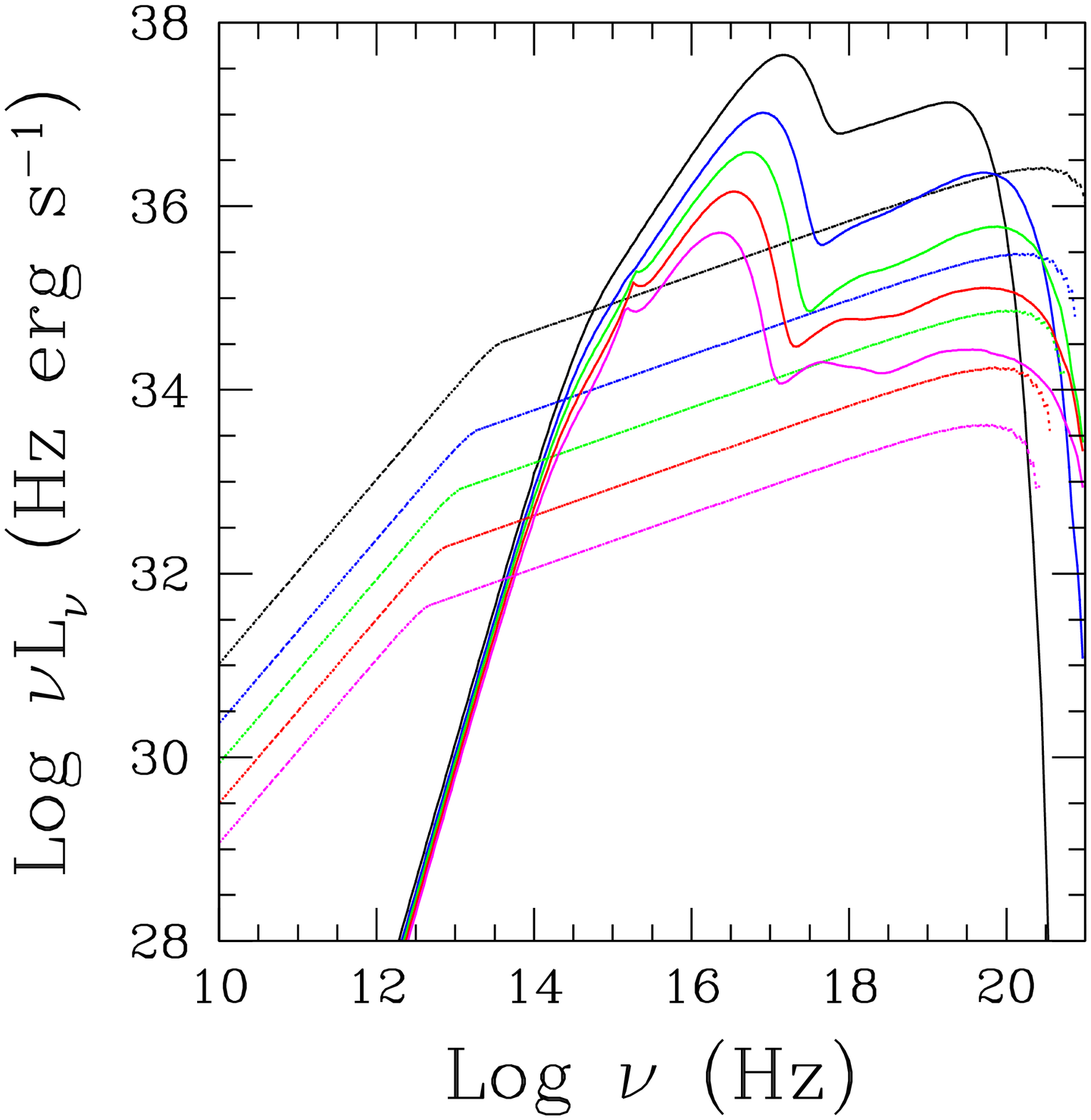} & 
\epsfxsize=5.5cm \epsfbox{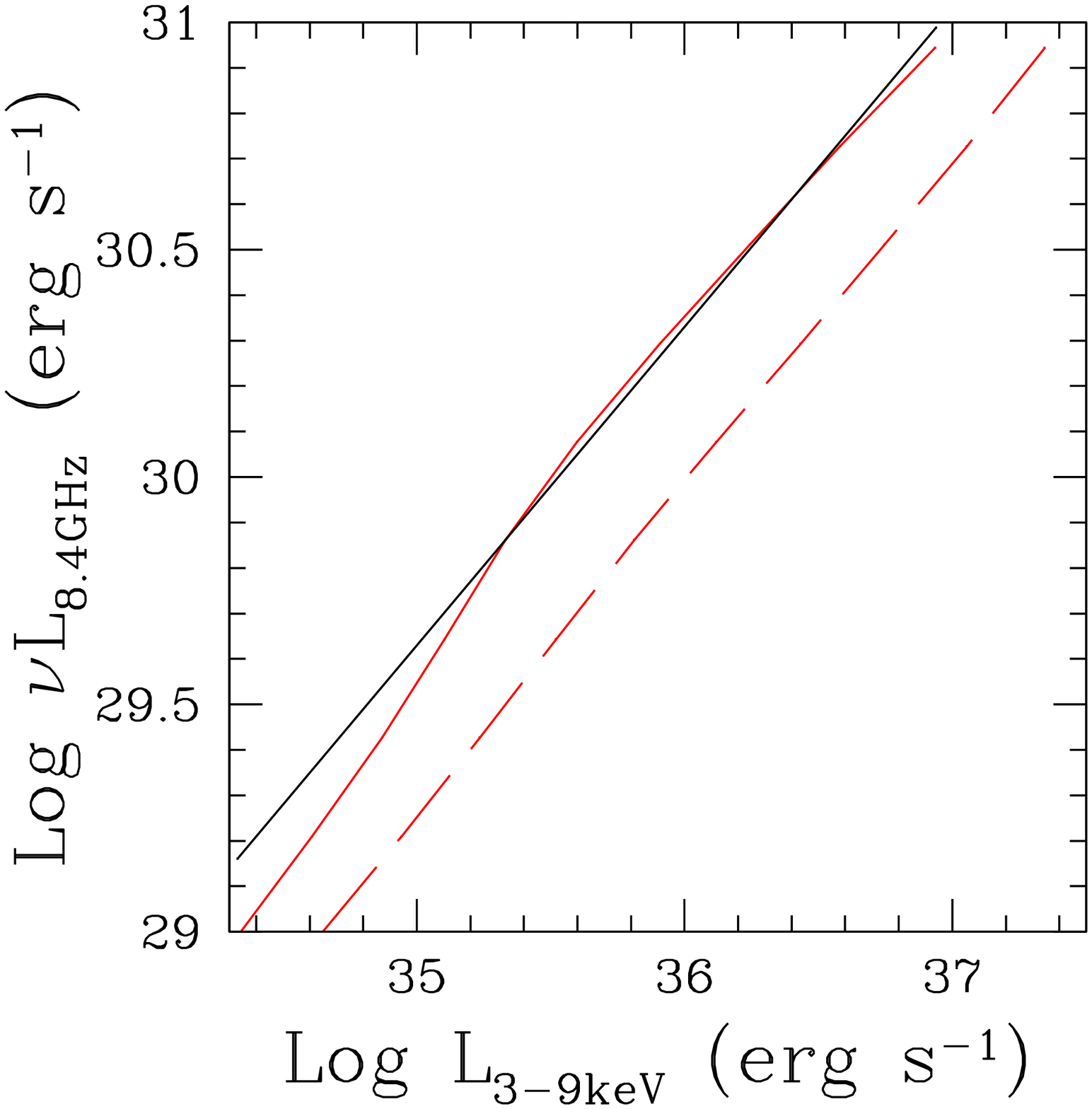} &
\epsfxsize=5.5cm \epsfbox{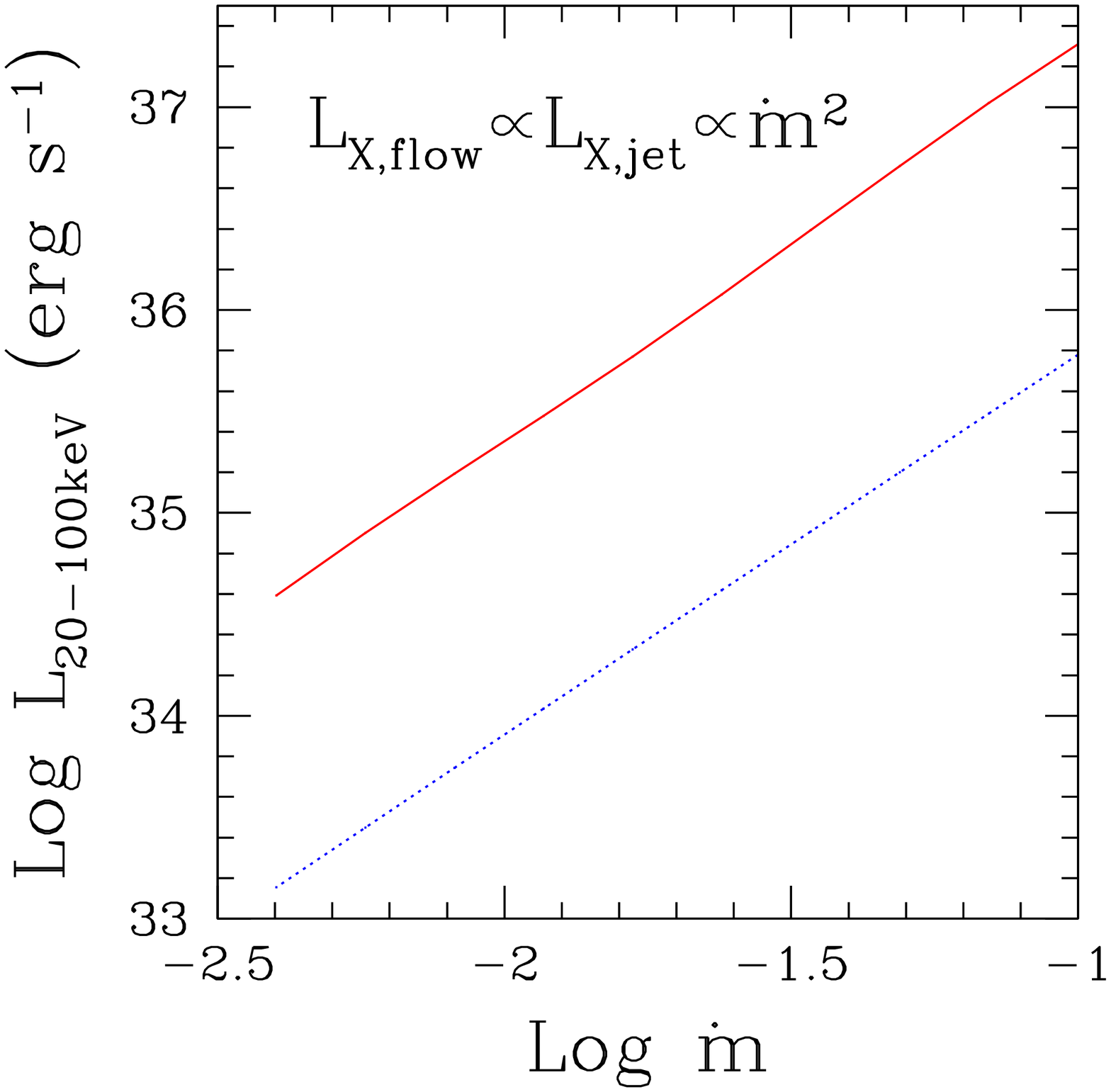}\\ 
\end{tabular}
\caption{a). Model SEDs, including synchrotron jet emission (dotted line) for increasing truncation radius: $20R_g$ (black), $35R_g$ (blue), $50R_g$ (green), $70R_g$ (red) and $100R_g$ (magenta). b). $L_R\propto L_X^{0.7}$ radio-X-ray correlation (black line), together with correlation from coupled accretion flow/jet model (solid red line). Dashed red line shows radio-X-ray correlation using $L_{20-100keV}$ X-ray luminosity instead of $L_{3-9keV}$. c). X-ray luminosity as a function of mass accretion rate, where $\dot{m}=\dot{M}/\dot{M}_E$, for X-rays from the radiatively inefficient accretion flow (solid red line) and X-rays from the jet (dotted blue line).}
\label{fig3}
\end{figure*}

\begin{table*}
\begin{tabular}{ccccccc}
\hline
Type of Jet & $L_R$ & $L_{X,jet}$ & $L_{X,flow}$ & Dominant Source of X-ray Emission & $\alpha_{flow}$ & $\alpha_{jet}$ \\
& & & & & ($L_R\propto L_{X,flow}^{\alpha}$) & ($L_R\propto L_{X,jet}^{\alpha}$)\\ 
(1) & (2) & (3) & (4) & (5) & (6) & (7) \\
\hline
Standard & $\dot{m}^{1.4}$ & $\dot{m}^2$ & $\dot{m}^2$ & Either Hot Flow or Jet & $0.7$ & $0.7$ \\
Cooling & $\dot{m}^{0.6}$ & $\dot{m}$ & $\dot{m}^2$ & Hot Flow with a transition to Jet X-rays as $\dot{m}$ decreases & $0.3$ & $0.6$ \\
\hline
\end{tabular}
\caption{Scalings with accretion rate for a standard self-similar conical synchrotron jet, and the same synchrotron jet including cooling. (2) Dependence of radio luminosity on Eddington scaled accretion rate. (3) Dependence of jet X-ray luminosity on accretion rate. (4) Dependence of flow X-ray luminosity on accretion rate, assuming a radiatively inefficient hot flow. (5) Dominant source of X-ray emission as a function of accretion rate. (6) Index of the radio-X-ray correlation for X-rays from the flow. (7) Index of the radio-X-ray correlation for X-rays from the jet.}
\label{table:scalings}
\end{table*}

The radio jet is an important part of the energy budget of the black
hole accretion flow, with kinetic energy comparable to the hard X-ray
luminosity at $\dot{m}_c$ (e.g. Cyg X-1: Gallo et al 2005; Russell et
al 2007; Malzac et al 2009). Hence we take the jet $L_{KE,max}= \dot{m}_c
L_{Edd}=1.3\times 10^{38}$ ergs s$^{-1}$. 

We add a standard conical jet model onto our accretion flow
(e.g. Blandford \& Konigl 1979; hereafter BK79; Merloni et al 2003;
Falcke et al 2004), assuming that some acceleration process operates
continuously down the jet, so that a small fraction of the electrons
in the jet form a relativistic particle distribution. The electrons radiate
via synchrotron to produce a broad band spectrum from radio to
X-rays. If these radiative losses are high then this will affect the
self similar jet structure. Hence we limit the radiative luminosity to
10 per cent of the kinetic luminosity of the jet i.e. we make a
maximally radiatively efficient, self similar jet.

Parameters for a standard conical jet include the distance from the
black hole at which the material is accelerated, $Z_0$ (the jet
base). We make the standard assumption that the energy is transported
by Poynting flux from $r_h$ in the hot flow, where the jet is
presumably launched, to $z_0$ without any radiative losses. The self
similar behaviour then extends out from $Z_0$ to a distance of
$Z_{max}=10^6Z_0$, where $Z=zR_g$ is distance along the jet. Distance
perpendicular to the jet is $R_j=\rho R_g=\phi Z$, where $\phi$ is
constant for a conical jet.

We use observations to set the bulk Lorentz factor $\Gamma=1.2$ and
opening angle $\phi=0.1$ (e.g. Gallo et al 2005). We assume that these
stay constant with $\dot{m}$. We transform all specific luminosities
$L_{\nu}$ from jet frame to observer frame by multiplying by
$\delta^3=(\Gamma-\sqrt{\Gamma^2-1}\cos\psi)^{-1}$, assuming a mean
inclination angle $\psi=60^\circ$, and boost all frequencies by
$\delta$.

In such a geometry, the magnetic field energy density $U_B(z) \propto
z^{-2}$ (BK79). Turbulence in the field probably results in scaling
between the relativistic particle and magnetic pressures, so
$U_{rel}(z)=m_ec^2 \int_1^{\gamma_{max}}N(z,\gamma) \gamma d\gamma =
f_{rel}U_B(z)$ where $N(z,\gamma)$ is the electron distribution at each
point $z$ of the jet. We make the standard assumptions that
$N(z,\gamma)=K(z) \gamma^{-p}$ with $p=2.4$ between $\gamma_{min}=1$
and $\gamma_{max}=10^5$.  Hence the optically thin synchrotron
emission has energy index $\alpha=(p-1)/2=0.7$ i.e. it rises in $\nu
f_\nu$ with energy output peaking at the highest frequency
$\nu_{max}=4/3\gamma^2_{max}\nu_B$.

The power law synchrotron emission becomes optically thick to
self-absorption below $\nu_{ssa}\propto K^{2/7}B^{5/7}R_j^{2/7}\propto
(z/z_0)^{-1}$ (Ghisellini et al 1985) i.e. decreases with larger
distance along the jet. The flux at this point, $L(\nu_{ssa})_{sync}
\propto (z/z_0)^{-1} (\nu_{ssa}/\nu_B)^{-(p-1)/2} dZ \propto
(z/z_0)^{-1} dZ \propto d \log Z$. Thus the self absorbed spectra from
each part of the jet sum together to produce the characteristic 'flat
spectrum' (i.e. energy index $\alpha=0$: BK79) at low frequencies.

We do not include the self consistent inverse Compton 
emission from the jet, since this contributes only at higher energies
(e.g. Zdziarski et al 2012) and the baseline model we are testing is
one where the X-rays are produced by synchrotron from the jet.

\subsection{Jet at $\dot{m}_c$}

We anchor the jet at $\dot{m}_c=0.1$ using observational constraints.
The observed break from optically thick to optically thin synchrotron
in a bright low/hard state from GX 339-4 is $\nu_{ssa,0}\sim
10^{13.5}$, and the 10GHz radio luminosity from the sum of self
absorbed jet components is six orders of magnitude below the X-ray
emission (Gandhi et al. 2011) i.e. $\nu L_\nu\sim 10^{31}$ ergs
s$^{-1}$ at 10GHz. This sets $B(z_0)\sim 3.5\times 10^4$~G
(i.e. $K(z_0)=2.4\times 10^{12}$ cm$^{-3}$ for $f_{rel}=0.1$) and
$z_0\sim 5300$. This gives a total radiated luminosity of 10 per cent
of the kinetic jet power, as described above for a maximal radiatively
efficient jet. 

We note that in these standard conical jets the fraction of radiative
power to kinetic power is not constant down the jet, as the radiation
depends on both magnetic and electron energy density so is $\propto
(z/z_0)^{-4} dV$ while the jet kinetic energy is simply $\propto
(z/z_0)^{-2} dV$.

\subsection{Jet scaling with mass accretion rate - transition to a jet 
dominated state?}

We then assume that the jet kinetic power $\propto \dot{m}$, and that
all energy densities scale as $\dot{m}/\dot{m}_c$ (Heinz \& Sunyaev
2003) so that $B(z,\dot{m})=B_0(z_0,\dot{m}_c) (z/z_0)^{-1}
(\dot{m}/\dot{m}_c)^{1/2}$ and $K(z,\dot{m})=K_0(z_0,\dot{m}_c)
(z/z_0)^{-2} (\dot{m}/\dot{m}_c)$. Fig 3a shows a sequence of spectra
for decreasing $\dot{m}$ using this coupled accretion flow-jet model.
Our model reproduces the $L_R\propto L_X^{0.7}$ radio-X-ray
correlation, as shown in Fig 3b. The X-rays come from a radiatively
inefficient accretion flow and are therefore proportional to
$\dot{m}^2$. The radio is from the optically thick jet, where it has a
flat spectrum so $L_R\propto B_0^{1.2}K_0^{0.8}\propto\dot{m}^{1.4}$
for any model where the magnetic and relativistic particle energy
densities scale with $\dot{m}$, hence $L_R\propto L_X^{0.7}$ for a
radiatively inefficient X-ray flow (e.g. Heinz \& Sunyaev 2003;
Merloni et al 2003). The model slightly deviates from this relation at
low $\dot{m}$, because of the spectral curvature in our model at low
$\dot{m}$ which changes the scaling over a small bandpass
($L_{3-9kev}$). The dashed line in Fig 3b shows that a wider bandpass
recovers the relation even down to the lowest luminosities. As
discussed in Section 2.1, this detailed issue can probably be
circumvented by a proper treatment of the self consistent electron
distribution in the hot flow (as in Veledina et al 2011).

The jet emission does not ever dominate the total hard X-ray emission,
but remains an approximately constant factor below the hot flow (Fig
3c). This is because the optically thin synchrotron luminosity
$L_{X,jet}\propto K_{0}U_{B,0}\propto\dot{m}^2$ so it also follows a
radiatively inefficient scaling. This is in contrast to the jet {\em
kinetic} luminosity, which does scale as $\dot{m}$. Thus while the
{\em kinetic} luminosity of the jet can easily dominate the radiative
energy of the flow, the {\em radiated} energy of the jet drops as fast
as that from the flow.  Thus there is no transition in the X-ray
spectrum from being dominated by the radiatively inefficient hot flow
to being dominated by the synchrotron emission of a conical,
self-similar jet (see also Merloni et al 2003; Falcke et al
2004). There are instances in the literature where there is a flow-jet
transition in the X-ray flux, but these use models where the flow is
radiatively efficient (no advection: Fender et al 2003) and/or have a
jet kinetic power which does not scale as $\dot{m}$ (Yuan \& Cui
2005).

\subsection{Jet scaling with electron cooling}

The discussion above assumes that the energy density in relativistic
particles scales as the energy density in the jet i.e. $\int
N(\gamma)\gamma d\gamma \propto \dot{m}/M$ (Heinz \& Sunyaev
2003). However, a better approach is to say that it is the injected
electron distribution, $Q(\gamma)=Q_0q(\gamma)$, which scales, and then cools into a
steady state distribution $N(\gamma)=Kn(\gamma)$ (e.g. for Blazar
jets: Ghisellini et al
2010). In this case, the injected distribution is normalised to the
available power (ie. $Q_0\propto\dot{m}$), so that $K$ now scales as
$K\propto Q_0/U_{seed}$. For synchrotron cooling,
$U_{seed}=U_B\propto\dot{m}$ giving $K$ constant with accretion
rate. Hence $L_{X,jet}\propto K_0U_{B,0}\propto\dot{m}$, such that it
is possible for the jet X-rays to overtake the X-ray luminosity from
the flow as accretion rate decreases. However, $L_R\propto
B_0^{1.2}K_0^{0.8}$, so the radio luminosity no longer scales as
$L_R\propto\dot{m}^{1.4}$ but scales as $L_R\propto\dot{m}^{0.6}$
(Table 1). When the X-rays come from the jet, this gives $L_R\propto
L_{X,jet}^{0.6}$, which is not inconsistent with the data. But for
higher accretion rates, when the X-rays come from the flow (and we
know $L_{X,flow}$ must be proportional to $\dot{m}^{\beta}$ where
$\beta>1$ for a transition to occur at all) this becomes $L_R\propto
L_{X,jet}^{0.3}$, which does not match the observed correlation. The
optically thick synchrotron from the jet must still drop as
$\dot{m}^{1.4}$ to make the $L_X-L_R$ relation when the hot flow
dominates.

\subsection{Composite Jet with electron cooling break}

Since energetic electrons cool faster, the electron distribution is
expected to be composite, with the electron distribution above some
break energy $\gamma_b$ being dominated by cooling, while below this
energy it reflects instead the injected electron distribution
(e.g. Markoff et al 2005; Yuan et al 2005; Zdziarski et al 2012). 
The cooling break energy $\propto 1/(U_B Z)\propto \dot{m}^{-1}$ 
(Zdziarski et al 2012, equation 36), so does not depend on mass but
increases linearly with decreasing mass accretion rate. For our
parameters, the cooling break is $\gamma_{cool}\sim 1.3$ for
$\dot{m}=\dot{m}_c$, increasing to  $\gamma_{cool}\sim 26$ for our
lowest $\dot{m}$. Even the highest  $\gamma_{cool}$ is mostly below
the synchrotron self-absorption break, so makes very little difference in the spectrum
(see e.g. Zdziarski et al 2012 fig 5a) or in the predicted X-ray 
scaling from the completely cooling dominated jet described above. 

\subsection{Arbitrary Jet scaling}

It is possible to contrive situations for a synchrotron jet where the
radio scales as $\dot{m}^{1.4}$ but the X-rays scale as $\dot{m}$
(eg. by allowing $z_0$ to change with accretion rate). But it is clear
any transition from a flow where $L_X\propto \dot{m}^2$ to this jet,
where $L_X\propto \dot{m}$, will cause a steepening of the observed
$L_R-L_X$ relation (Yuan \& Cui 2005). Changing the X-ray behaviour
with $\dot{m}$, without a simultaneous change in the behaviour of the
optically thick radio emission, necessarily changes the $L_R-L_X$
correlation in a way which is not observed (Corbel et al 2013). Since
we do not observe a change in the radio-X-ray correlation down to
quiescence in BHBs (e.g Corbel et al 2013 and references therein), we know
there can be no transition in dominant X-ray production mechanism down
to these luminosities. Whatever dominates the X-rays in the brightest
LHS spectra, at the top of the correlation, must dominate at the
bottom. This rules out all plausible models in which the X-rays switch
from being dominated by the flow to being dominated by the jet.

We note that a break has been observed in the radio-X-ray correlation in AGN at very low accretion rates ($\dot{m}\sim 10^{-6}$, Yuan, Yu \& Ho (2009)), but crucially the observed minimum in photon index occurs where the correlation is unbroken, implying that jet emission taking over cannot be the cause. 

\subsection{Jet dominated models}

Since a switch from flow dominated to jet dominated X-ray flux is
ruled out by the radio-X-ray correlation, the final possibility is
that the X-rays are always dominated by the jet (Falcke et al 2004).
However our fiducial jet model is already very efficient at producing
radiation. To make the jet dominate at $\dot{m}_c$ would require that
almost all of the jet kinetic energy was transformed to radiation,
which seems unlikely. It would also impact on our assumptions that
adiabatic and radiative losses are negligible, and leaves unanswered
the question of what causes the change in X-ray behaviour (hardening
then softening of the spectral index) as $\dot{m}$ decreases. 
Jet dominated models are also unable to fit the high energy rollover
seen at high energies seen in the bright low/hard state
($\dot{m}\sim \dot{m}_c$: Ibragimov et al 2005; Torii et al 2011). 
More spectral and energetic constraints are discussed in Zdziarski et al
(2003) and Malzac et al (2009). 

\section{Conclusions}

The observed change in X-ray spectral index as $\dot{m}$ decreases in
the low/hard state (first hardening then softening) can be
quantitatively explained by a truncated disc/radiatively inefficient
hot inner flow. Seed photons from the disc drop as the disc recedes
with decreasing $\dot{m}$ so that self generated cyclo-synchrotron
seed photons in the flow become dominant in Compton cooling. 
This model can also produce the observed radio-X-ray correlation
with the addition of a standard, conical self-similar jet. 
These standard jets are as radiatively inefficient as the hot
flow, so there is no transition from the X-rays being dominated by the
flow to being dominated by the jet, which was the alternative
explanation for the X-ray spectral behaviour (Russell et al 2010,
S11). Including the effects of cooling allows the jet X-rays to drop more slowly with accretion rate and hence overtake the X-rays from the hot flow, however such a transition would also necessarily distort the radio-X-ray
correlation in a way which is not observed.

Thus we show that the truncated disc/radiatively inefficient hot
flow/standard conical jet model can quantitatively explain the broad
band spectral evolution of BHB in the low/hard state, with the X-rays
always being dominated by the flow, and the radio by the jet. However,
at low luminosities, the optical depth in the hot flow is rather
small, so the X-ray spectra are no longer a power law and the
individual Compton scattering orders can clearly be seen. Yet the
observed BHB spectra at these low mass accretion rates (quiescence)
are well described by a power law (e.g. Gallo et al 2006). This
discrepancy is even more evident in the low mass accretion rate AGN
(e.g Yu, Yuan \& Ho 2011), which has again led to models where the
X-rays are dominated by the jet. Since these are inconsistent with the
observed radio-X-ray correlation, we suggest instead that this points
to a more complex flow, where the electron acceleration process
produces an intrinsically non-thermal distribution. Thermalisation via
cyclo-synchrotron emission and absorption produces the dominant
thermal electron population of the bright low/hard state, while the
dramatic increase in seed photons from the disc in the high/soft state
means that the power law distribution is seen (Malzac \& Belmont 2009;
Vurm \& Poutanen 2009). We suggest at very low accretion rates the cooling is so inefficient that thermalisation does not happen and the electron distribution remains non-thermal. A non-thermal electron distribution will emit a power law spectrum, as observed. Hybrid thermal/non-thermal models, especially
combined with a multi-zone approach (Veledina et al 2012), hold out
the possibility to understand the broad band spectral variability both
on long timescales, with changing accretion rate, and on short timescales,
to understand how fluctuations in the flow can make the observed 
IR/optical/X-ray correlations (Kanbach et al 2001; Malzac et al 2004;
Ghandi et al 2008).

\appendix 
\section{Accretion Flow Model}

The model consists of an outer black body disc (BB), truncated at some radius ($R_{t}$), with an inner hot flow of radius $R_{hot}$, where $R_{hot}=20R_g$. The hot flow is taken to be radiatively inefficient, such that: 

\begin{equation}
L_{hot}=L_{BBdisc}(R<R_{t})\left(\frac{\dot{m}}{\dot{m}_c}\right)
\end{equation}

Where $\dot{m}=\dot{M}/\dot{M}_E$, $\dot{m}_c=0.1$ and $L_{BBdisc}(R<R_{t})$ is the luminosity of a BB disc extending from the truncation radius down to the last stable orbit. 

We scale the truncation radius with accretion rate, such that:

\begin{equation}
\dot{R_{t}}=20R_g\left(\frac{\dot{m}}{\dot{m}_c}\right)^{-1/2}
\end{equation}

The optical depth ($\tau$) of the Comptonising region is fixed at 2 for $\dot{m}=\dot{m}_c$, and scales with $\dot{m}$ as:

\begin{equation}
\tau=2\left(\frac{\dot{m}}{\dot{m}_c}\right)
\end{equation}

The unabsorbed cyclo-synchrotron emission from the hot flow is calculated following Di Matteo et al. (1997):

\begin{equation}
L_{cyclo}(\nu)=5.57\times 10^{-29}\frac{n\nu I(x)V}{K_2(1/\theta_e)}
\end{equation}

Where $V=2/3\pi R_{hot}^3$ is the volume of the Comptonising hot flow, $n$ is the number density of electrons calculated from the optical depth, $\theta=kT/m_ec^2$ is the dimensionless electron temperature, $K_2(1/\theta)$ is the modified Bessel function, $x=2\nu/3\nu_B\theta^2$, $\nu_B=eB/2\pi m_ec$ is the Larmor frequency, and the function $I(x)$ is given by:

\begin{equation}
I(x) = \frac{4.050}{x^{1/6}}\left(1+\frac{0.40}{x^{1/4}}+\frac{0.532}{x^{1/2}}\right)\exp(-1.8899x^{1/3})
\end{equation}

The magnetic field ($B$) of the hot flow is calculated from the density by assuming the ions are at the virial temperature and the magnetic field is $10\%$ of the gas pressure, giving:

\begin{equation}
B = \sqrt{0.1nm_pc^2 \frac{8\pi}{r_{hot}}}
\end{equation}

Where $r_{hot}=R_{hot}/R_g$. The cyclo-synchrotron self-absorption frequency is given by:

\begin{equation}
\nu_{csa}=\frac{3}{2}\nu_B \theta^2 x_m
\end{equation}

Where $x_m$ is found by solving for $x$ when the cyclo-synchrotron and BB emission are set equal:

\begin{equation}
L_{cyclo}(\nu_{csa}) = 8\pi^2m_e\nu_{csa}^2\theta_e R_{hot}^2
\end{equation}

Below the self-absorption frequency the absorbed emission is calculated as:

\begin{equation}
L(\nu<\nu_{csa})=\left(\frac{\nu}{\nu_{csa}}\right)^{5/2}L(\nu_{csa})
\end{equation}

Thermal Comptonisation is modelled using {\sc eqpair} (Coppi 2002), with seed photons from both the disc and cyclo-synchrotron emission, where the fraction of disc photons from a given radius ($R$) intercepted by the hot flow is given by:

\begin{equation}
\frac{L_{seed,disc}}{L_{disc}}=\left(\frac{R_{hot}}{R}\right)\frac{\arcsin(R_{hot}/R)}{\pi}
\end{equation}

And we calculate the mean seed photon temperature:

\begin{equation}
kT_{seed}=\frac{k(L_{seed,disc}T_{disc}+L_{cyclo}T_{cyclo})}{L_{seed,disc}+L_{cyclo}}
\end{equation}

The electron temperature, a parameter in both the cyclo-synchrotron equations and {\sc eqpair}, is calculated self consistently. 

\subsection{Jet Model}

We construct a conical jet, where opening angle ($\phi=0.1$) relates jet radius ($R_j$) to distance along the jet ($z$):

\begin{equation}
R_j(z)=\phi z
\end{equation}

We assume a fraction of the accreting material ($f_j$) is diverted up the jet. The energy density in relativistic particles at the jet base is set to be some fraction ($f_{rel}=0.1$) of the magnetic energy density:

\begin{equation}
m_ec^2\int_{\gamma_{min}}^{\gamma_{max}}\gamma n(\gamma)d\gamma=U_{rel,0}= f_{rel}U_{B,0}
\end{equation}

We conserve magnetic energy and particle number along the jet such that:

\begin{equation}
B(z)=B_0\left(\frac{z}{z_0}\right)^{-1}
\end{equation}
\begin{equation}
K(z)=K_0\left(\frac{z}{z_0}\right)^{-2}
\end{equation}

And allow $B_0$ and $K_0$ to scale with accretion rate as:

\begin{equation}
B_0=B_0(\dot{m}_c)\left(\frac{\dot{m}}{\dot{m}_c}\right)^{1/2}
\end{equation}
\begin{equation}
K_0=K_0(\dot{m}_c)\left(\frac{\dot{m}}{\dot{m}_c}\right)
\end{equation}

Where $\dot{m}_c=0.1$, and $B_0(\dot{m}_c)$, $K_0(\dot{m}_c)$ and $z_0$ are fixed by requiring the radio luminosity and the optically thick-optically thin synchrotron break match observations of GX 339-4 (Gandhi et al. 2011). 

We assume electrons in the jet are continually accelerated into a power law distribution of the form:

\begin{equation}
n(\gamma)=K\gamma^{-p}
\end{equation}

Where $p=2.4$, for electron Lorentz factors ranging from $\gamma=1.0-1\times10^5$. 

We split the jet into conical sections and calculate the synchrotron emission from electrons in each section:

\begin{equation}
L_s(\nu)=\frac{\sigma_Tc}{8\pi\nu_B}U_B\gamma n(\gamma)V\delta^3
\end{equation}

Where $V$ is the volume of the conic section, $\delta=1/(\Gamma-\cos{\psi}\sqrt{\Gamma^2-1})$ is the boosting factor of the jet, $\psi$ is the angle of the jet with respect to the observer, and the electron Lorentz factor and synchrotron photon frequency are related by $\gamma=\sqrt{3\nu/4\nu_B}$. 

The synchrotron self-absorption frequency ($\nu_{ssa}$) in each section is given by (Ghisellini et al. 1985):

\begin{equation}
\nu_{ssa}=\left(4.62\times10^{14}KB^{2.5}\frac{R_j}{0.7}\right)^{2/7}
\end{equation}

The frequency of the observed radiation is boosted by a factor $\nu_{obs}=\nu\delta$. We neglect synchrotron self-Comptonisation. 

\label{lastpage}

\end{document}